\documentclass{vgtc}                          

\ifpdf
  \pdfoutput=1\relax                   
  \usepackage{graphicx}                
  \DeclareGraphicsExtensions{.pdf,.png,.jpg,.jpeg} 
\else
  \usepackage{graphicx}                
  \DeclareGraphicsExtensions{.eps}     
\fi%

\graphicspath{{figures/}{pictures/}{images/}{data/}{./}}

\usepackage{microtype}                 
\PassOptionsToPackage{warn}{textcomp}  
\usepackage{textcomp}                  
\usepackage{mathptmx}                  
\usepackage{times}                     
\usepackage{cite}                      
\usepackage{tabu}                      
\usepackage{booktabs}                  

\usepackage{lipsum}
\usepackage{layouts}
\usepackage{gensymb}
\usepackage{textcomp}
\usepackage{microtype}

\usepackage{siunitx}
\newcommand{\singlequote}[1]{\lq{#1}\rq{}}

\onlineid{0}

\vgtccategory{Research}

\vgtcinsertpkg

\title{Visual Analytics for Early Detection of Retinal Diseases}

\author{Martin Röhlig\thanks{e-mail: martin.roehlig@uni-rostock.de}\\ %
\parbox{1.5in}{\scriptsize \centering Visual and Analytic Computing \\ University of Rostock} %
\and Oliver Stachs\thanks{e-mail: oliver.stachs@uni-rostock.de}\\ %
\parbox{1.7in}{\scriptsize \centering Department of Ophthalmology \\ Rostock University Medical Center} %
\and Heidrun Schumann\thanks{e-mail: heidrun.schumann@uni-rostock.de}\\ %
\parbox{1.5in}{\scriptsize \centering Visual and Analytic Computing \\ University of Rostock}}

\teaser{
\centering
\includegraphics[width=\textwidth]{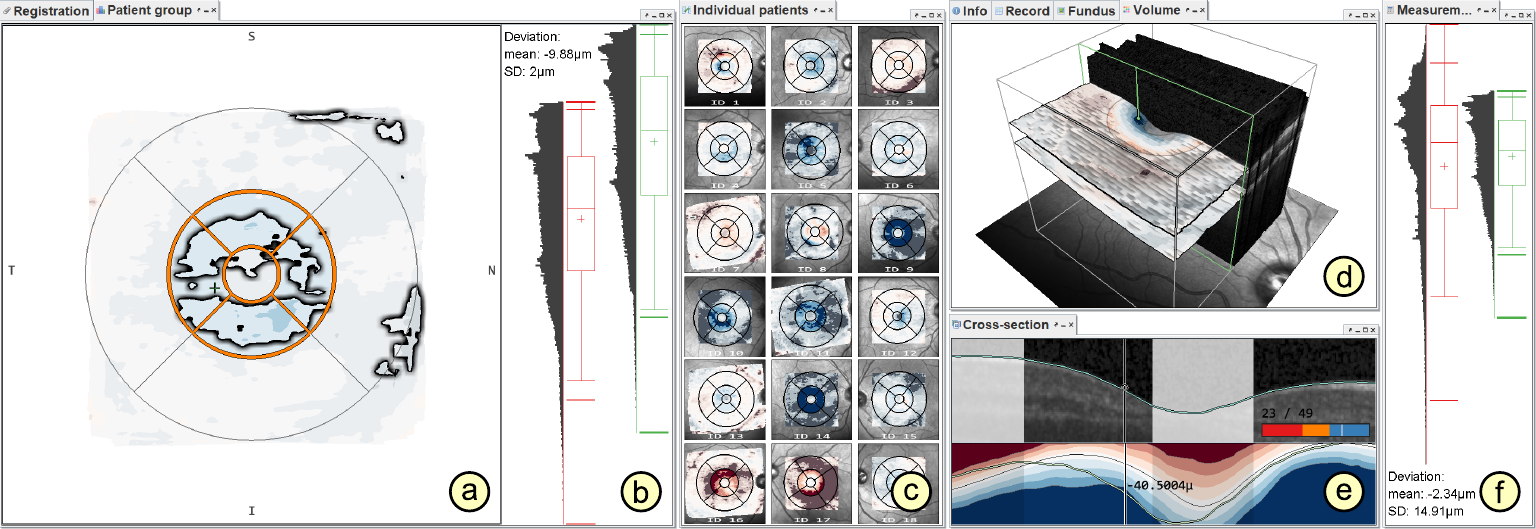}
\caption{%
Interactive exploration of early retinal changes in single and multiple patient OCT data.
Shown is the comparison of retinal layer data of a patient group with a control group via a deviation map (a), measurements of a selected map region (b), and deviation maps for all individual patients (c).
The 3d image data of a single selected patient is visualized as a volume (d), a cross-section through the volume (e), and patient-specific measurements for the region selected in the group map (f).
}
\label{fig:teaser}
}

\abstract{
Advances in optical coherence tomography (OCT) have enabled noninvasive imaging of substructures of the human retina with high spatial resolution.
OCT examinations are now a standard procedure in clinics and an integral part of ophthalmic research.
The interpretation of the OCT helps ophthalmologists understand the impact of various retinal and systemic diseases on the structure of the retina in a way not previously possible.
In the early stages of retinal diseases, however, the identification and analysis of small and localized substructural changes in the retina remains a challenge.
We present an overview of novel visual analytics approaches for the interactive exploration of early retinal changes in single and multiple patients, the comparison of the changes with normative data, and automated quantification and measurement of diagnosis-relevant information.
We developed these approaches in close collaboration with ophthalmology researchers and industry experts from a leading OCT device manufacturer.
As a result, they not only significantly reduced the time and effort required for OCT data analysis, especially in the context of cross-sectional studies, but have also led to several new discoveries published in biomedical journals.
} 

\CCScatlist{
\CCScatTwelve{Human-centered computing}{Visu\-al\-iza\-tion}{Visu\-al\-iza\-tion application domains}{Visual analytics}
}

\usepackage[nameinlink]{cleveref}

\crefname{part}{Pt.}{Pts.}
\Crefname{part}{Part}{Parts}
\crefname{chapter}{Ch.}{Chs.}
\Crefname{chapter}{Chapter}{Chapters}
\crefname{section}{Sect.}{Sects.}
\Crefname{section}{Section}{Sections}
\crefname{subsection}{Sect.}{Sects.}
\Crefname{subsection}{Section}{Sections}
\crefname{subsubsection}{Sect.}{Sects.}
\Crefname{subsubsection}{Section}{Sections}
\crefname{figure}{Fig.}{Figs.}
\Crefname{figure}{Figure}{Figures}
\crefname{table}{Tab.}{Tabs.}
\Crefname{table}{Table}{Tables}

\begin{document}


\firstsection{Introduction}\label{sec:introduction}

\maketitle

Advances in optical coherence tomography (OCT)~\cite{HSLS91} enable the noninvasive in vivo acquisition of 3D images of the human retina with ever higher resolution, quality, and accuracy.
Ophthalmologists interpret the OCT to understand a variety of retinal and systemic disorders.
These include common ocular diseases, such as age-related macular degeneration, diabetic retinopathy, or glaucoma~\cite{KPLH12,VMCH15,GrTa13}, as well as other pathologies with ocular signs, such as multiple sclerosis~\cite{DWBG11}.
Hence, the number of clinical OCT examinations has steadily increased worldwide, in parallel with a continuous growth of OCT-based research in ophthalmology~\cite{FuSw16}.
This progress in OCT image acquisition and availability is accompanied by increasing efforts to effectively process and analyze the data.

Modern OCT scanners capture hundreds of cross-sections of retinal tissue with micrometer resolution in just a few seconds~\cite{DMGK01}.
Using the images, ophthalmologists can differentiate the fine-grained substructures of the retina -- \emph{the retinal layers} -- which helps them to assess photoreceptor defects, abnormal axonal thickness, or macular degeneration~\cite{YoHa14}.
However, the complexity of the data makes it difficult to translate the growing number of 3D images into insights.

\begin{figure*}
\centering
\includegraphics[width=\textwidth]{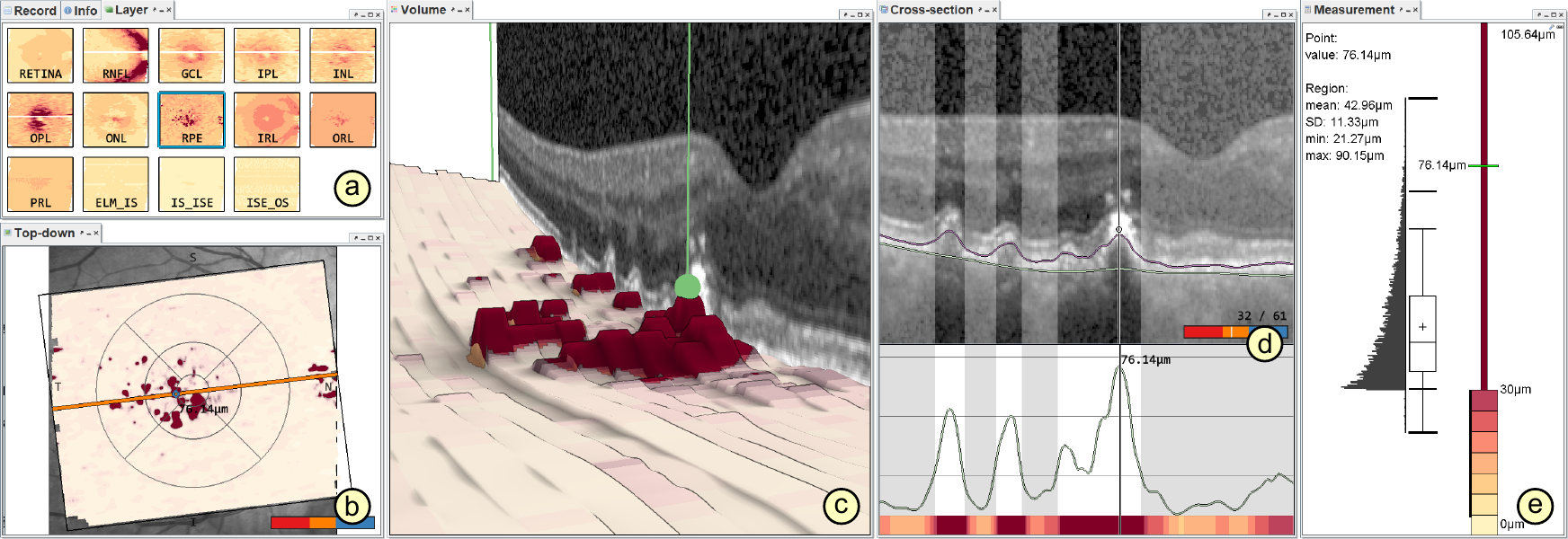}
\caption{\label{fig:approach:maps}
VA of a single patient with four linked views: top-down view~(a, b), 3D view~(c), cross-sectional view~(d), and measurement view~(e).
Selected regions with high attribute values (dark red) are highlighted in the different views.
}
\end{figure*}

The general goal is to identify exactly the information within the overall data that is relevant to the diagnosis of a patient or the evaluation of a research study.
At the early stages of a disease, this is especially ambitious.
The reason is that initially often only local areas of certain layers are affected by small changes in the structure of the retina.
Where these changes occur, however, is not known in advance.
For single datasets, each included cross-section has to be examined one by one, and for multiple datasets, simplifications need to be applied to reduce the amount of data.
Both current methods involve a high manual analysis effort and the risk of losing valuable information.
Nevertheless, for the early detection of retinal diseases, it is crucial to find out which layers are affected and to what extent.

From a visualization perspective, the 3D data cannot be easily displayed in a single picture without obscuring the retinal layers.
This hinders the visibility of small and local defects.
As more and more image data are acquired with modern OCT devices, such details become even harder to see.
The data complexity also makes it difficult to select relevant parts within the 3D data for detailed analysis.
Thus, new strategies for detecting early signs of ocular disease in retinal OCT data are needed.

To this end, we developed novel visual analytics (VA) approaches for the interactive exploration of retinal changes in single and multiple patient OCT data.
Partial aspects of our work have already been separately published in VA-related and biomedical journals~\cite{RSPR18,RPSS19,PRFG19,PMFS20}.
Here, we present a summarizing overview of these individual parts of our ongoing VA research. 
Generally, our VA methods focus on the computation of diagnostically relevant attributes of retinal layers and their interactive visualization with high spatial specificity.
They were developed in close collaboration with ophthalmologists and industry experts.
Together, we designed an effective combination of visual and statistical analysis methods, alongside an improved procedure that considerably reduces the analysis effort compared to the currently applied analysis approach.
As a result, our methods have enabled ophthalmologists to gain new insights into the early effects of diseases on the structure of the retina in two cross-sectional studies.
We are working to improve our solutions in current follow-up studies and further planned studies.

\section{Background}\label{sec:background}

The sensitivity of modern OCT scanners allows the different retinal layers to be differentiated and precisely measured, close to an in-vivo \singlequote{optical biopsy} of the retina~\cite{AdDu13}.
The commercial system used in our work (Spectralis\textsuperscript{\textregistered} OCT, Heidelberg Engineering GmbH, Heidelberg, Germany) is based on frequency domain optical coherence technology.
It captures cross-sections of the retina at high speed (axial scan rate of \SI{85}{\kilo\Hz}) with an axial and lateral resolution of \SI{3.5}{\micro\metre\per pixel} and \SIrange{6}{11}{\micro\metre\per pixel}.
Multiple cross-sections are combined into 3D images and further processed, including noise reduction and segmentation of 11 retinal layers~\cite{EWFP14}.
The resulting data helps ophthalmologists to assess changes in the retinal structure that cannot be detected by other examination methods.

For the early detection of retinal changes, patient- and group-specific evaluations of single and multiple OCT datasets are performed.
Data sizes can vary widely depending on the type and objective of the studies.
In our collaborations, the smallest dataset included 19 cross-sections with \SI{512 x 496}{pixels}, while the largest had 241 cross-sections with \SI{1024 x 496}{pixels}.
A higher resolution and number of cross-sections are required to detect local defects and minimize measurement artifacts~\cite{ODLS11}.
Yet, the increased complexity also complicates the data exploration, which becomes more difficult when comparing multiple datasets.

Related VA research has mostly focused on visualization of single OCT datasets~\cite{GAWR09,GBMJ14,GKFZ09}.
The early detection of retinal changes and ophthalmic studies have hardly been considered so far.
This is why mainly commercial OCT software is used to analyze OCT-based study data, e.g., from Zeiss or Heidelberg Engineering.
For single datasets, it allows to flip through cross-sections to inspect and measure the retinal layers.
With many cross-sections, this can be cumbersome and time-consuming.
Multiple datasets are instead reduced based on ETDRS grids~\cite{ETDR91_Grading}.
These grids divide the retina into 9 regions for which average measurements are statistically analyzed.
This provides an overview, but risks information loss due to aggregation of local changes.
In our studies, the ETDRS grids compressed the data to only about \SI{0.0034}{\percent} of the values per slice.

\section{Visual analytics of retinal layers}\label{sec:approach}

Our VA research aims to provide automated computation, adequate visual representation, and interactive exploration for early retinal defects.
Showing the OCT images alone or applying general simplifications, however, makes it difficult to detect small and local changes of retinal layers.
We therefore focus on the extraction of diagnosis-relevant layer attributes, their visualization in linked 2D and 3D views, and measurements of regions that deviate from normative data.
This way, the data are selectively reduced, while preserving early changes in retinal structure.
Compared to the currently applied approach, all relevant parts are directly visible and accessible.

Thanks to a flexible design, our methods allow ophthalmologists to focus on relevant information not only in single patient OCT data, but also in multiple patient data.
As described in \Cref{sec:application}, a complementary analysis procedure additionally reduces the effort for data analysis and enables new insights in ophthalmic studies.

\subsection{Visual analysis of individual datasets}\label{sec:approach:maps}

The goal of analyzing individual OCT datasets is to determine which part of a single patient's retina is affected by a disease.
Especially with early retinal defects, this is hardly possible by visualizing only cross-sections of the 3D OCT images.
Our idea therefore is to reduce the 3D layers in the images to 2D attribute maps~\cite{RSPR18}.
These maps capture different properties of the layers and help to focus the analysis on information that is relevant for a diagnosis.

\paragraph*{Computation of layer attributes:}

We designed several novel attribute maps in order to detect various retinal defects caused by ocular diseases.
A number of layer attributes have been incorporated, including existing attributes, e.g., layer thickness~\cite{ZSMZ96}, and new ones, e.g., layer curvature.
The attribute values are calculated point by point and compiled into attribute maps.
This allows us to accurately characterize even local changes in the structure of the retina.
During visualization, the layers can then be analyzed by interactively browsing the 2D maps.
This reduces the layers' complexity and clarifies their different properties in 2D and 3D views of the OCT data.

\paragraph*{Visualization of layer attributes:}
When visualizing retinal layers, it is important that no local defects are overlooked in the 3D spatial context.
Our new visualization design allows to start from an overview of all layers and then selectively narrow down the data to the layer parts of interest.
Coordinated selection and highlighting helps associated attribute values to be inspected in detail.

As shown in \Cref{fig:approach:maps}, our design consists of four linked views.
A top-down view provides an overview of the colored attribute maps of all layers in a dataset (\cref{fig:approach:maps}a), with one selected layer enlarged over an image of the inner eye surface for spatial reference (\cref{fig:approach:maps}b).
A 3D view highlights the 3D spatial structure of the selected layer in a blended volume visualization together with the OCT image data (\cref{fig:approach:maps}c).
A cross-sectional view allows detailed inspection of layer boundaries and associated attribute values in a line chart (\cref{fig:approach:maps}d).
A measurement view statistically summarizes the attributes of selected layer regions (\cref{fig:approach:maps}e), e.g., with particularly high values (dark red).

Using the views together helps the user to browse the layers, switch between attributes, and identify those with relevant characteristics.
Interesting layer parts can then be directly accessed, their structure analyzed in 2D and 3D, and interactively measured.
This improves the visibility of relevant information and simplifies the analysis compared to the currently applied approach, where only one attribute and individual cross-sections can be viewed at a time.

\begin{figure}
\centering
\includegraphics[width=\columnwidth]{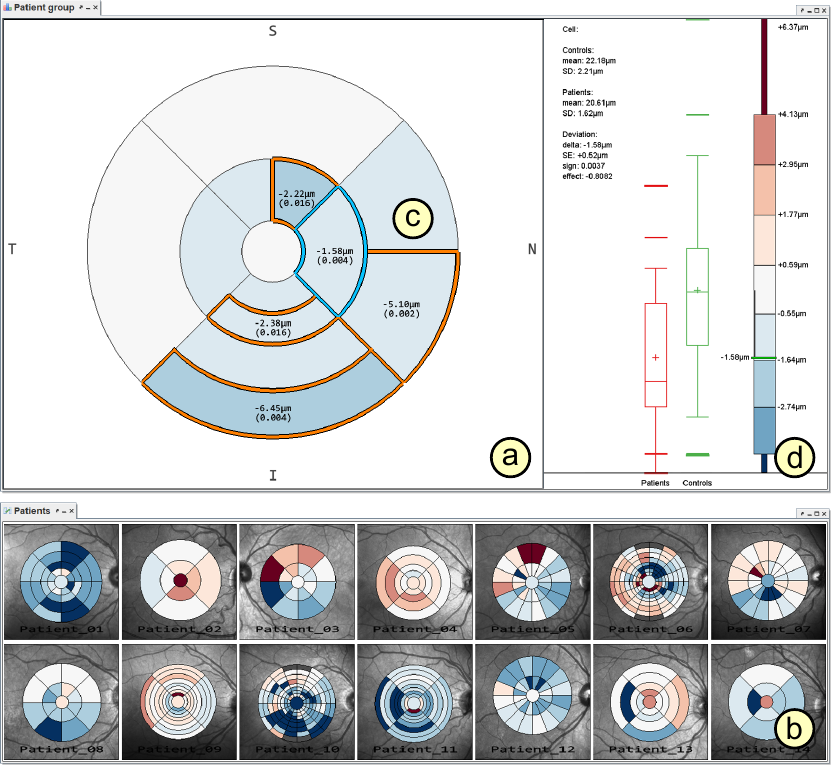}
\caption{\label{fig:approach:grids}
VA of a patient group~(a) and individual group members~(b) relative to control data.
Cells with significant differences (orange cell border) are highlighted~(c) and summarized~(d).}
\end{figure}

\subsection{Visual analysis of multiple datasets}

Multiple OCT datasets from a number of patients are analyzed to determine common patterns of retinal changes and how these changes differ from normative data.
This requires a stronger abstraction of the retinal layers, but also takes local changes into account.
Hence, we introduced adaptive grids that selectively reduce the data while preserving local variations in attribute values~\cite{RPSS19}.
In addition to the attribute maps, these grids allow statistical and visual comparison of data from patient groups with data from control groups.

\paragraph*{Computation of adaptive grids:}
The existing ETDRS grid divides the retina into anatomically important areas, but provides only a fixed and coarse abstraction.
In our design, we extend the ETDRS layout by incrementally subdividing its grid cells.
For each cell, the average and a summary of all enclosed attribute values are computed.
To obtain dataset-specific grids, the variations of the values in the cells are evaluated.
Cells with low variation are well represented by the average values, while high variation indicates information loss.
On that basis, we suggest the best fit for given data with the highest acceptable variation and lowest cell count.
This is also done for multiple datasets by specifying a common grid layout that represents the average attributes of a patient or control group.

\paragraph*{Visualization and comparison of grids and maps:}
For the visualization, we must take care that the variations in attribute values captured by the adaptive grids can be explored in the spatial context.
Hence, we integrate the grids in our four linked views (\cref{sec:approach:maps}).
The top-down view shows a suitable grid for each layer colored and labeled by the average cell values.
To retrieve more or less detail, the cells can be interactively refined by splitting or merging, or replaced by sections of the attribute maps.
The measurement and other linked views provide further information about the values within the cells.

On top of that, our design supports the comparison of multiple individual patients or patient groups with control data (\cref{fig:approach:grids}).
For this, deviations per grid cell or map point from control intervals are encoded with a diverging color palette.
To quantify the observed differences, we apply statistical tests to every grid cell or map point.
The resulting areas with significant differences are highlighted in the linked views (\cref{fig:approach:grids}c).
The measurement view provides further statistical information, e.g., $p$-values and effect sizes, for selected cells or map regions (\cref{fig:approach:grids}d).
This way, abnormal layer regions of single patients, aggregated patient groups (\cref{fig:approach:grids}a), and individual group members (\cref{fig:approach:grids}b) can be assessed with high spatial specificity.

Our combined visual and statistical analysis helps to focus on relevant information specific to single and multiple patients.
Here, the data are not rigidly reduced, but to the level that fits the given layers.
In contrast, the currently applied approach relies only on statistical analysis of ETDRS grids without spatial context.

\subsection{Evaluation and feedback}

We developed our designs with ophthalmologists and experts from a leading OCT device manufacturer (Heidelberg Engineering GmbH, Germany).
Together, we devised and tested our solutions using a participatory design and pair analytics~\cite{AKGF11} approach.
We also collected structured feedback through interviews and questionnaires.
From the ophthalmologists' point of view, the previous problems in the analysis of early retinal defects could be addressed.
Especially with regard to ophthalmic research, they noted that our work can be an essential component in developing new biomarkers.
Our methods have thus been applied in several studies (\cref{sec:application}).
From the industry experts' perspective, our solutions are an important addition to the existing method range.
A future embedding of our solutions in commercial OCT software was considered highly desirable.

\section{Application and results}\label{sec:application}

Our VA approaches deliver new visualization techniques and new interactive analysis methods for retinal OCT data in ophthalmology.
A particular focus of ophthalmic research has been the question how primary diseases, e.g., diabetes mellitus, affect the retina, even when signs of secondary ocular complications, e.g., diabetic retinopathy, have not been clinically detected.
In our efforts to answer this question, we first presented our VA methods and initial results to the scientific community in ophthalmology in two articles~\cite{RJFP17,SJSR17} and at major conferences~\cite{PRFJ18,SPFS19}.
Together with ophthalmologists, we then complemented our methods with a new analysis procedure and applied it in two cross-sectional studies.
The detailed results proved that our methods not only reduced the analysis effort, but also enabled new discoveries.

\paragraph*{New analysis procedure with reduced effort:}

\begin{figure}
\centering
\includegraphics[width=\columnwidth]{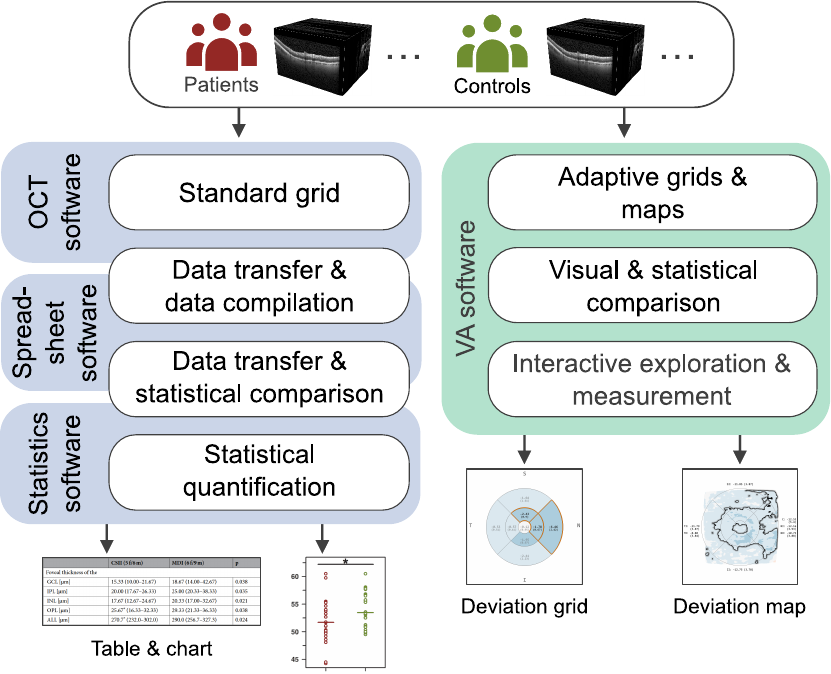}
\caption{\label{fig:approach:workflow}
Comparison of the steps of the currently used analysis procedure and our analysis procedure with deviation maps and grids.}
\end{figure}

Our analysis procedure focuses on single- and multi-patient data in cross-sectional studies~\cite{RPSS19}.
To assess its performance, we first evaluated two studies~\cite{GKPJ18,PGKJ18} with the currently applied approach and then used our new VA methods on the study data (\cref{fig:approach:workflow}).
The comparison of both approaches confirmed that the currently applied mix of commercial and non-commercial software requires more time and effort for the data analysis.
This is mainly because switching software and transferring data between analysis steps must be done manually.
We also found that the statistical analysis of ETDRS grids actually did not capture all local layer defects and that the used tables and charts made it difficult to relate the results to the specific areas of the retina.
In contrast, with our new VA procedure, we were able to provide extended support for most analysis steps in the study evaluation.
That allowed us to generate results faster and with less effort, while preserving local variations and their spatial context.

\paragraph*{Cross-sectional studies with new findings:}

\begin{figure}
\centering
\includegraphics[width=\columnwidth]{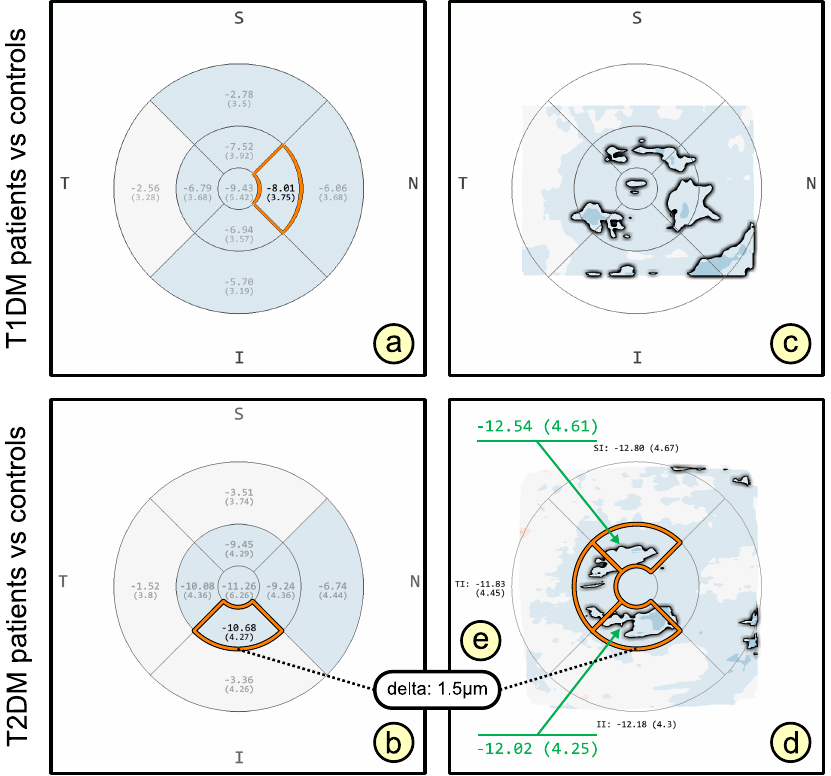}
\caption{\label{fig:application:study}
Comparison of patients with type 1 and type 2 diabetes mellitus to controls.
The ETDRS grids (a and b) show less significant differences (orange cell borders) than the maps (c and d; black outlines).
The grids also underestimate the differences (e), while the maps allow to measure irregularly shaped regions (green arrows).
}
\end{figure}

We evaluated our new VA methods in two cross-sectional studies investigating the early effects of diabetes mellitus on retinal structure.
Besides the primary disease, no signs of secondary diabetic retinopathy were detected in all subjects by other examination methods.
Still, as shown in \Cref{fig:application:study}, our VA methods provided additional insights into the diseases' influence compared to the currently applied approach.

The first study focused on identifying early changes in retinal thickness in children with type 1 diabetes mellitus~\cite{PRFG19}.
Our new attribute maps revealed more local areas of significant differences between the patient and control groups than the currently applied ETDRS grid-based analysis (\cref{fig:application:study}a, c).
For the first time, the ophthalmologists were also able to observe the exact spatial extent of the differences in the maps, even across cell boundaries.
The second study then concentrated on measuring mild to moderate changes in retinal thickness in adults with type 2 diabetes mellitus~\cite{PMFS20}.
The results showed that the currently applied ETDRS grids tend to underestimate differences due to averaging of cell values (\cref{fig:application:study}e).
With our new maps, however, the irregularly shaped areas with significant differences could be interpreted as single defects and measured more accurately.
Similar accurate results were also obtained with our new adaptive grids in both studies, but with the added benefit of an adjusted spatial resolution~\cite{RPSS19}.

Our combination of OCT and VA methods thus led to new insights into exactly which retinal layers are affected where and to what degree by early changes in the diseases studied.
Follow-up and further studies with our methods are planned.

\section{Summary and outlook}\label{sec:summary}

We summarized our ongoing VA research on the interactive exploration of retinal substructures in single and multiple patient OCT data.
Thanks to the computation of diagnosis-relevant attributes, their visualization as maps and adaptive grids with high spatial specificity, and the interactive measurement of abnormal regions, ophthalmologists were able to identify and assess even small and local substructural defects.
A complementary analysis procedure helped to significantly reduce the effort required for OCT data analysis, especially for cross-sectional studies.
As a result, new insights were gained into the influence of primary diseases -- diabetes mellitus type 1 and 2 -- on the retinal structure.
This indicates that our work contributes to the early detection of retinal diseases in a way not possible before.

Besides the designs presented here, we introduced several extensions, including a unified access to retinal OCT data from different sources~\cite{RSPR18}, an advanced measurement method for the comparison of grids and maps~\cite{RSPS19}, a VA method for parameter influences on OCT data~\cite{RLPS17}, and a new compound map to display all layer attributes in a dataset at once~\cite{SRSP18}.
In our continued cooperation with ophthalmologists, we are now considering other data aspects in combination with these extensions to further analyze the early defects that can be identified with our solutions.
One aspect is time to capture the dynamics of structural changes in current follow-up and other planned longitudinal studies.
Early results in breast cancer patients suggest that our methods are effective in detecting temporal differences in retinal thickness~\cite{Stache2022,Stache23}.
However, more work is needed to allow comparisons of more than two time points simultaneously.
Another aspect are other examination methods, e.g., OCT angiography or visual field measurements, to investigate structural and functional changes of the retina together.
These research directions will be helpful in monitoring disease progression, response to treatment, and biomarker development in different patient groups.


\bibliographystyle{abbrv-doi}

\bibliography{bibliography}
\end{document}